\documentstyle [psfig,supertab] {l-aa}  
 
\def\teff{\ifmmode T_\star \else $T_\star$\fi}
\def\logg{\ifmmode \log g \else $\log g$\fi}

\begin{document}

\thesaurus{08(09.08.1;  09.01.1; 08.01.3; 08.13.2; 08.05.1; 13.25.5;)}
 
\title{Combined stellar structure and atmosphere models for massive stars}
\subtitle{IV. The impact on the ionization structure of single star H~{\sc ii} regions}

\author{Gra\.{z}yna Stasi\'{n}ska\inst{1} \and Daniel Schaerer\inst{2}}
\offprints{G. Stasi\'{n}ska}

\institute{DAEC, Observatoire de Meudon, 92195 Meudon Cedex, France
(grazyna@obspm.fr) \and 
Space Telescope Science Institute, 3700 San Martin Drive, Baltimore, 
MD 21218, USA (schaerer@stsci.edu)}

\date{Received 6 August 1996 / Accepted October 1996} 
\maketitle 
\markboth{G. Stasi\'{n}ska \& D. Schaerer: The impact of {\em CoStar} 
  models on single star H~{\sc ii} regions}
   {G. Stasi\'{n}ska \& D. Schaerer: The impact of {\em CoStar} models
  on single star H~{\sc ii} regions} 

\begin{abstract} 
We study the impact of modern stellar atmospheres that take into 
account the effects of stellar winds, departures from LTE and line 
blanketing (``{\em CoStar}'' models) on the ionization structure of 
H~{\sc ii} regions. 
Results from a large grid of photoionization models are presented.
Due to a flatter energy distribution in the He~{\sc i} continuum, 
compared to the widely used Kurucz models, generally higher ionic 
ratios are obtained.

We find that N$^{+}$/O$^{+}$ and Ne$^{++}$/O$^{++}$ can be safely 
used as direct indicators of N/O and Ne/O abundance ratios in 
H~{\sc ii} regions, over a wide range of astrophysical situations.

The roughly constant observed value of Ne$^{++}$/O$^{++}$ ionic ratios 
in Galactic H~{\sc ii} regions is naturally reproduced by photoionization
models using {\em CoStar} fluxes, while Kurucz models at solar metallicity fail 
to reproduce this behaviour. 
This gives support to ionizing fluxes from non--LTE atmospheres including
stellar winds and line blanketing.
However, we also point out that tests of stellar atmosphere models from 
observations of H~{\sc ii} regions are hampered by a lack of strong
constraints on the ionization parameter.


\end{abstract}

\section{Introduction} Most of the recent work on H~{\sc ii} regions
modelling uses
the Kurucz (1991) plane-parallel LTE model atmospheres. There are two reasons to
this. One is that Kurucz's computations include line blanketing, which is indeed
important in stellar atmospheres, especially at high metallicities. This
effect was
not included in earlier non--LTE stellar atmosphere modelling  (e.g. Mihalas
 \& Auer
1970).   The other reason is that Kurucz models are available for a wide set of
stellar temperatures and metallicities, which makes them extremely useful
for any
study aimed at reproducing the observed properties of H~{\sc ii} regions in
different environments. 

However, it is recognized that Kurucz models cannot predict correctly the
distribution of the radiation in the Lyman continuum of hot stars, not only
because
non-LTE effects are neglected, but also because the presence of winds in
such stars
is expected to have an impact on the outcoming stellar flux (Gabler et al.~1989,
Najarro et al.~1996, Schaerer et al.~1996b).

A new set of theoretical stellar spectra from the combined stellar structure and
atmosphere models (referred to as {\em CoStar} models) of Schaerer et al.~(1996a, 
b) has now been published (Schaerer \& de Koter  1996, hereafter SdK96). The
atmosphere, in particular, includes the  effects of stellar winds, departure
from
LTE as well as blanketing.  It is the purpose of this paper to explore the
impact
of these new  ionizing fluxes on single star H~{\sc ii} region models. 

In Section 2, we recall the main features of the {\em CoStar} models, and
discuss the expected uncertainties. In Section 3, we present a grid of 
photoionization models constructed for metallicities solar (Z\sun) and one 
tenth solar (Z\sun/10), using the {\em CoStar} models and the Kurucz ones, 
and discuss the main differences. In Section 4, we present some implications 
on H~{\sc ii} regions diagnostics. 
The possibility of testing these new stellar atmosphere models using 
observations of H~{\sc ii} regions is discussed in Sect.~5. Section 6 
summarizes the main conclusions of this study.

\section{Comparison of {\em CoStar} and Kurucz model atmospheres}

\subsection{Comparison of ionizing fluxes} A detailed comparison of the ionizing
fluxes from {\em CoStar} models with  the plane parallel LTE models of
Kurucz and
also recent non--LTE  models of Kunze et al.~(1992) and Kunze (1994) is given in
SdK96. Here we shall only briefly recall the most important differences
between the
{\em CoStar} and the Kurucz fluxes used in our study.
Wind models similar to ours have also been presented by Sellmaier et 
al.~(1996).

As an illustration we show in Fig.~1 a comparison between the {\em CoStar} and 
Kurucz model fluxes for two dwarf models with $T_\star$ $\simeq$ 46000 and 
35000 K respectively. Also indicated are the positions of some
ionization potentials relevant to the discussion in forthcoming sections. 
For the {\em CoStar} model we plot the continous spectral energy distribution
with the spectral resolution used in the non-LTE calculation. Blanketing due to 
numerous spectral lines of all elements up to zinc is included in these models 
by means of an opacity sampling technique (using 20 $\rm\AA$ bins) as described 
in Schaerer \& Schmutz (1994). The ionization edges of metals (mostly CNO, Ne, 
and Ar) found in other models (e.g.~Kunze et al.~1992, Kunze 1994) are not 
treated in our 
models. On the other hand these may be overestimated if iron peak elements
causing an important fraction of metal line blanketing are not included.
We note that the recent models of Sellmaier et al.~(1996) also lack of 
pronounced metal ionization edges in the EUV, which seems to confirm our
calculations. As discussed in SdK96, a more detailed analysis of this 
question will be possible with the inclusion of additional metals in the 
full non-LTE calculations (see e.g.~de Koter et al.~1994, 1996).

The largest differences between the Kurucz and {\em CoStar} models appear in 
the He~{\sc ii} continuum above 54 eV: due to the
velocity gradient in the stellar wind a strong non--LTE effect depopulates the
He~{\sc ii} groundstate and leads to a decrease of the bound-free opacity
above 54 eV (cf.~Gabler et al.~1989). This explains the larger 
flux in the He~{\sc ii} continuum in non--LTE models accounting for stellar 
winds ({\em CoStar}) compared to the Kurucz models.
Of more importance for the present study is the flux distribution at  lower
energies, where the ionization potentials of the most important  elements
observed in H~{\sc ii} regions are located. The essential characteristics 
of the {\em CoStar} spectrum in the He~{\sc i} continuum (24.6 eV $< E <$ 
54 eV) are (cf.~SdK96): {\em 1)} a flatter spectral energy distribution, and  
{\em 2)} an increase of the total flux. 
The flattening of the spectrum mainly results from a
wind effect  whose strength decreases towards higher energies, since the 
continuum is formed at larger optical depths. 
In the considered temperature range non--LTE effects increase the total 
He~{\sc i} ionizing flux in both plane parallel models and those accounting 
for stellar winds.

\subsection{Adopted CoStar and Kurucz models}

The {\em CoStar} models are available for two sets of abundances:  
Solar (Z=0.020 in the notation of SdK96) and low metallicity (Z=0.004). 
We will compare these models with the Kurucz ones at solar and 
one tenth solar metallicity, which is the closest possible correspondance
in metallicity.

In the following, we will compare {\em CoStar} model atmospheres for main
sequence stars (the ones labeled A2--F2 in SdK96) with the available 
Kurucz models having the closest effective temperature and gravity. 
We will also consider {\em CoStar} models for giant stars (the ones 
labeled A4--D4 in SdK96). 

The {\em CoStar} models provide not only the spectral distribution of the
ionizing radiation, but also the stellar radii.  The radii of stars 
modelled with the Kurucz atmospheres are obtained from the calibration of 
Vacca et al.~(1996) for stars of luminosity class V and Ia, for main sequence
and giants respectively.

\section{Photoionization models}

\subsection{Definition of the grid} The photoionization models are
constructed with
the code PHOTO, the latest description of which can be found in Stasi\'{n}ska \&
Leitherer (1996, hereafter SL96). A model computed with this version of PHOTO
that can be directly compared with the results obtained with other
photoionization
codes (presented at the Lexington workshop, Ferland et al.~1995) is also
given in
SL96 \footnote{Extensive results (intensities of about 100 lines in the
optical, UV
and IR domains, mean ionic fractions and averaged electron temperatures) for the
photoionization models computed for this study are accessible by anonymous ftp
on ftp.obspm.fr in the directory /pub/obs/grazyna/SS96 and most of them can be
found as well on the CD-ROM accompanying the paper ``A Database for Galaxy
Evolution Modeling'', by Leitherer et al.~(1996).}.

The ionization structure of a nebula is mainly determined by the ionizing radiation
field and by the nebular geometry. In the simple case of 
 a sphere of constant gas density $n$ and constant filling factor  $\epsilon$, the
ionization structure of the heavy elements --- for a given spectral distribution 
of the ionizing radiation field --- is essentially a function of the ionization
parameter $U=Q_{\rm H}/(4 \pi R^2 \, n \, c)$, where $Q_{\rm H}$ is the number of 
stellar photons emitted per second above 13.6  eV, $R$ is the external radius, 
and $c$  is the speed of light.
It can easily be shown that any combination of  $Q_{\rm H}$, $n$ and $\epsilon$
that keeps $Q_{\rm H} \, n \, \epsilon^2$ constant results in a similar 
ionization structure. For each star, we have thus computed a series of 4 
models of hydrogen density $n=$ 10 $\rm cm^{-3}$ and $\epsilon$=1, in which 
the total stellar luminosity has been multiplied  by a factor 
$10^6$, $10^3$, 1, $10^{-3}$ respectively for the models numbered from 1 to 4. 
Doing this, one covers the entire range of expected ionization parameters, 
from the most compact H~{\sc ii} regions (number 1) to the most diluted 
interstellar medium (number 4). Numbers 2--3 correspond roughly to the range 
of ionization parameters encountered in giant H~{\sc ii} regions as found 
in SL96. The ionization structure of model number 2 is the same as that of 
a model with a single ionizing star and a gas density of
$n = 10^4 \, {\rm cm^{-3}}$. From now on, models number 2 will be called the
``reference models'' \footnote{Actually, a sphere may not be the best
approximation to the geometry of H~{\sc ii} regions. Therefore, we have also 
introduced a parameter $f$, which is the internal radius of the nebula, 
in units of the Str\"omgren radius for a full sphere at an electron temperature 
of 10000 K. We have computed series of models with $f = 10^{-2}$ and $f = 3$. 
In the latter case, which is  close to a plane parallel model, the average
``excitation'', as measured for example by [O~{\sc iii}]/[O~{\sc ii}] tends 
to be lower than for a full sphere of
the same $U$ (i.e.~with $Q_{\rm H}$, $n$, $f$ and $\epsilon$ combining into the
same value of $Q_{\rm H} \, n \, \epsilon^2 \, (1+f^3)^{-2}$). This effect is
especially important for low effective temperature stars. In this paper, we
present only results for $f = 10^{-2}$, but the interested  reader will find 
the $f = 3$ models on the anonymous ftp account.}.
 
Whithin the range of chemical compositions expected in  H~{\sc ii} regions, 
the ionization structure of a nebula is only little dependent of its actual 
chemical composition. 

In building our grid, the chemical composition of the gas was taken solar
(O/H= $8.51 \, 10^{-4}$ with the abundances of the remaining elements as in
SL96) when dealing with solar metallicity stellar atmospheres, and one tenth
solar(O/H=$8.51\, 10^{-5}$) when dealing with low metallicity atmospheres. 

\subsection{Comparison of photoionization models using {\em CoStar} and Kurucz 
fluxes} The comparison is made in a series of diagrams showing ratios of ionic
fractions $x({\rm X^{+n}})/x({\rm Y^{+m}})$ as a function of the
stellar effective temperature $T_\star$ \footnote{$x({\rm X^{+n}})$ is 
defined as: $\int{n({\rm X^{+n}}) n_e dV}/ \int{n(\rm X)/ n_e dV}$)}.
Figures 2 and 3 represent the solar abundance models using the {\em CoStar} 
fluxes and the Kurucz atmospheres respectively. 
Figures 4 and 5 are equivalent to Figs.~2 and 3 respectively,
but for the low metallicity models. In each figure, the ``reference models'' with
different $T_\star$ are  connected with a thick continuous line for main
sequence stars, and a thick dotted line for giants. For each star, models with
different values  of $U$ are connected with a thin dotted line.  

In Figures 2 -- 5 panels a -- c give examples of ratios of ionic fractions
that are supposed to indicate the stellar effective temperature (e.g.~Rubin 
et al.~1995). Panel a shows $x{(\rm He^{+}})/x{(\rm H{^+}})$, which depends 
on the ratio of stellar photons above 24.6 eV and 13.6 eV. 
Panel b shows $x({\rm N^{++}})/x({\rm N^{+}})$, and Panel c shows 
$x({\rm S^{+++}})/x({\rm S^{++}})$ for the solar abundance models and 
$x({\rm Ar^{+++}})/x({\rm Ar^{++}})$ for the low metallicity ones.  
While the $x({\rm He^{+}})/x({\rm H^{+}})$ is accessible from
optical and radio measurements of H~{\sc ii} regions (assuming a 
canonical helium abundance), $x({\rm N^{++}})/x({\rm N^{+}})$ and 
$x({\rm S^{+++}})/x({\rm S^{++}})$ are accessible from infra-red 
measurements.  $x({\rm Ar^{+++}})/x({\rm Ar^{++}})$ can
be obtained from optical lines, but only when the electron temperature is
sufficiently high for the lines to be measurable, that is to say only at low
metallicities. Other ratios involving two subsequent ionization stages of
the same element that can be compared in the same way with the observations 
are $x({\rm O^{++}})/x({\rm O^{+}})$,  $x({\rm Ne^{++}})/x({\rm Ne^{+}})$, 
or $x({\rm S^{++}})/x({\rm S^{+}})$.

The fact that at a given $T_\star$, the {\em CoStar} models have higher 
fluxes in the $\sim$ 30 eV - 54 eV region translates into higher ionic ratios in
photoionization models computed with {\em CoStar}, as can be seen in Panels
a -- c of Figs.~2 -- 5.
 This is conspicuous in ratios involving ions of high ionization potential like
S$^{++}$ (34.83 eV) or Ar$^{++}$ (40.74 eV). On the other hand,  the effect is
actually marginal on the $x({\rm He^{+}})/x({\rm H^{+}})$ ratio for main
sequence stars, being noticeable only at effective temperatures smaller 
than $\sim$ 40000 K. The reason is that, at higher temperatures, there are enough 
photons above 24.6 eV to fully ionize helium, so that 
$x({\rm He^{+}})/x({\rm H^{+}})$ becomes independent of the model atmosphere.  

An effective temperature determined from an ionic ratio such as the ones mentioned
above (assuming that one has some idea of the ionization parameter)  would thus be
{\em lower} using the {\em CoStar} models than using the Kurucz  models. Note that
a similar trend was found by Rubin et al.~(1995) when they compared
photoionization models constructed using the non-LTE atmospheres of Kunze (1994)
rather than the Kurucz ones. If indeed one were to use nebular models to determine
the effective temperature, one  would of course also have to take care that the
volume sampled by the  observations are comparable to the volume over which the
mean ionic  fractions are computed in the models.
Furthermore one should also remember that the atomic data used in the 
photoionization model computations are still incomplete: dielectronic 
recombination is generally an important process in the ionization balance, 
but the coefficients for ions of sulfur and argon are not available. Also,
charge-transfer reaction rates are only approximately known for many ions. 
In spite of these difficulties, it may be worthwhile noting that {\em CoStar} 
models predict a detectable [Ar~{\sc iv}] 4740 line in a wider range of 
effective temperatures than the Kurucz models.

Another important issue is that with {\em CoStar} we have appropriate 
models for all the evolutionary stages of massive stars. 
Models for giants tend to produce a higher excitation of the 
surrounding H~{\sc ii} regions than what is expected from extrapolating 
the dwarf models towards lower effective temperatures. 
This is clearly seen in Figs. 2 and 4 where the curves representing the 
giants conspicuously depart from the curves corresponding to main sequence 
stars. The giant model with \teff $\simeq$ 32000 K 
is a very strong case, where wind effects flatten the spectrum in the 
20 -- 40 eV range (cf.~SdK96).
When using the available Kurucz models, the only difference between 
giants and main sequence stars is in the total luminosity, which induces 
only a negligible variation in the excitation of the surrounding nebulae, 
through an increase of the ionization parameter $U$.  

\section{Some consequences for abundance determinations in H~{\sc ii} 
regions}

Empirical abundance determinations in H~{\sc ii} regions generally use an
ionization correction factor scheme first proposed by Peimbert \& Costero 
(1969) based on the similarity of ionization potentials of several ions. 
For example, to
derive the N/O ratio, one generally assumes that  N/O=N$^{+}$/O$^{+}$, and to
derive the Ne/O ratio, one assumes that  Ne/O=Ne$^{++}$/O$^{++}$. Actually,
photoionization models do not always  agree with these empirical schemes
(e.g.~Rubin 1985, Stasi\'{n}ska 1990).  This is not surprising, since the
ionization potentials of O$^+$ and N$^+$  or Ne$^{+}$ are not exactly the 
same, and the various recombination processes do not have the same strength 
for these ions. Rubin et al.~(1988) have strongly questioned the N/O ratios 
derived from optical measurements of H~{\sc ii} regions (partly because 
of the implied ionization correction scheme), and they consider the N/O
ratios derived from far infrared line measurements assuming
N/O=N$^{++}$/O$^{++}$ to be far more reliable. 

Panels d and e in Figs. 2 -- 5 show $x({\rm N^{+}})/x({\rm O^{+}})$ and 
$x({\rm N^{++}})/x({\rm O^{++}})$ respectively as a function of $T_\star$. 
We see that, actually, our models show the N$^+$/O$^+$ to be an excellent 
approximation for N/O in all the range of effective temperatures considered, 
except for the Kurucz models at solar metallicity. 
In this case, the slope of the stellar spectrum aroud 30 eV
is much larger than for the other models, inducing a differential effect between
N$^{+}$ and O$^{+}$. Thus, if {\em CoStar} models describe the stellar radiation
field more accurately than the Kurucz ones, the widely used empirical
N/O=N$^{+}$/O$^{+}$ scheme remains a quite safe approximation. Determining
the N/O ratio using N$^{++}$/O$^{++}$, on the other hand, gives rise to 
important biases since, at high effective temperatures, N$^{++}$ partly 
transforms into N$^{+++}$ and at low effective temperature
(when actually N$^+$ and O$^+$ are the dominant ionic species), the predicted 
$x(\rm N^{++})/x(\rm O^{++})$ becomes much larger than unity.

Panel f in Figs. 2 -- 5 plots $x({\rm Ne^{++}})/x({\rm O^{++}})$  as a function of
$T_\star$. We see that, in all the ``reference models'', this ratio decreases with
decreasing effective temperature, whether one uses the {\em CoStar} atmospheres or
the Kurucz ones. However, the range in $T_\star$ where $x({\rm
Ne^{++}})/x({\rm O^{++}})$ is close to unity is much larger when using the {\em
CoStar} atmospheres, because they have flatter spectral energy distributions in
the 30 -- 40 eV range than the Kurucz  ones (see also next Section).  Thus, if
{\em CoStar} predicts correctly the stellar energy distribution in  the Lyman
continuum, one expects the Ne$^{++}$/O$^{++}$ to give directly  the Ne/O ratios
over a wide range of parameter space relevant to H~{\sc ii} regions.

\section{Possible observational tests of stellar atmosphere models  
using H~{\sc ii} regions}

Since the predictions on the ionization structure of H~{\sc ii} regions 
depend so strongly on the type of model atmosphere, one might hope to use 
observations of H~{\sc ii} regions to test the flux of these models in 
the Lyman continuum. 

Along this line, it has already been noted that the so-called 
[Ne~{\sc iii}] problem, which appeared when using the Kurucz model 
atmospheres, seemed resolved when using model atmospheres, which, 
similar to {\em CoStar} models, account for stellar winds, and include 
a detailed non--LTE treatment and line blocking (Sellmaier et al.~1996, 
hereafter SYPR96). This [Ne~{\sc iii}] problem was made especially clear when
Simpson et al.~(1995; cf.~references therein) obtained far infrared line 
measurements for a sample of
H~{\sc ii} regions and plotted the values of Ne$^{++}$/O$^{++}$ as a function of
O$^{++}$/S$^{++}$.  They noted that the observations 
indicated an essentially flat  Ne$^{++}$/O$^{++}$ in contradiction
with the predictions of photoionization models using the Kurucz atmospheres 
(assuming no systematic variation in Ne/O or S/O).

In Figures 6 -- 9, we show the predictions of our model HII regions 
for $\rm Ne^{++}/\rm O^{++}$ to be compared directly to observations.  
Figures 6 and 7 represent the solar abundance models using the {\em CoStar} 
fluxes and the Kurucz atmospheres respectively. 
Figures 8 and 9 are equivalent to Figs.~6 and 7 respectively, but for the 
low metallicity models. The layout of the figures is
similar to the one used in Figs.~2 -- 5. Panel a plots
${\rm Ne^{++}}/{\rm O^{++}} = {\rm Ne/O} \, x({\rm Ne^{++}})/x({\rm O^{++}})$ 
as a function of 
${\rm O^{++}}/ {\rm S^{++}} = {\rm O/S} \, x({\rm O^{++}})/x({\rm S^{++}})$. 
Here, Ne/O and O/S are the standard abundance ratios used by Simpson et 
al.~(1995) (Ne/O=0.2 and O/S=47). In Figure 10, we plot the values of 
$\rm Ne^{++}/\rm O^{++}$ versus $\rm O^{++}/\rm S^{++}$ derived from 
infrared line ratios for the HII regions observed by Simpson et al.~(1995), 
using the atomic data referenced in SL96, and discarding
measurements with upper limits only.  Our models using {\em CoStar} have a
similar behaviour to the ones of SYPR96), and are in much better agreement 
with the observations than models based on Kurucz atmospheres. Actually, 
for the hottest stars, we obtain slightly larger values of  
$x({\rm O^{++}})/x({\rm S^{++}})$ at a given temperature $T_\star$ than SYPR96. 
Indeed, the {\em CoStar} atmospheres seem to produce somewhat higher fluxes 
in the 40 -- 54 eV energy range than the atmospheres of SYPR96. 
Detailed comparisons will be undertaken to understand the origin of these
differences.  
It appears, however, difficult to distinguish between both cases from nebular
observations.

Of course, one should bear in mind that different photoionization codes may
produce slightly different ionization structures. A direct comparison can 
be made for the models using the Kurucz atmospheres. Our computations using 
PHOTO seem roughly in agreement with the ones of SYPR96, as can be judged by 
comparing our Fig.~7 with their Fig.~2, and are slightly different from the 
ones presented in Simpson et al.~(1995).

One can also perform a similar observational test of the model atmospheres by
using optical lines, and plotting Ne$^{++}$/O$^{++}$ versus O$^{++}$/O$^{+}$,
as shown in panel b of Figs.~6 -- 9. This plot has the advantage of being
independent of S/O. Figure 11 shows the observed values for a sample of Galactic
H~{\sc ii} regions (from Shaver et al.~1983, filled circles) and of
Magellanic Cloud H~{\sc ii} regions (from Pagel et al.~1978, open circles), 
derived using the atomic data listed in SL96. The trend seen in 
Fig.~11 is more compatible with the {\em CoStar} models than with the Kurucz 
ones at solar metallicity. However, it must be noted that with 0.25 solar 
metallicity models (more relevant to Magellanic Cloud nebulae), the problem 
with the Kurucz models is less severe than at solar metallicity.

It is worthwhile recalling, that, in their abundance study of low metallicity
H~{\sc ii} galaxies from optical spectroscopy, Masegosa et al.~(1994) noted a
completely different [Ne~{\sc iii}] problem.  Their sample showed a slight
tendency for the observed Ne$^{++}$/O$^{++}$ to increase as O$^{++}$/O$^+$ 
{\em decreases} or as the H$\beta$ equivalent width decreases. 
If Ne$^{++}$/O$^{++}$ were to measure the abundance ratio Ne/O, this would 
mean that Ne/O depends on the evolutionary stage of the H~{\sc ii} galaxy. 
This is of course not acceptable, and Masegosa et al.~suspected that the 
Ne$^{++}$/O$^{++}$ ratio was affected by charge exchange. 
Our models do take charge exchange into account. The fact that the charge
transfer reaction O$^{++}$ + H$^0$ is far more efficient than the 
Ne$^{++}$  + H$^0$ one produces a differential effect between oxygen and 
neon ions, which becomes more conspicuous as the proportion of neutral 
hydrogen becomes greater in the O$^{++}$ region. 
This is seen in our models (Figs.~6 -- 9): the Ne$^{++}$/O$^{++}$ ratio
increases when the ionization parameter decreases.  The effect found by
Masegosa et al.~(1994) is illustrated in Fig.~12, where we plot 
Ne$^{++}$/O$^{++}$ versus O$^{++}$/O$^+$ as derived in the sample of low 
metallicity H~{\sc ii} galaxies studied by SL96 (which was based mainly 
on the Masegosa et al.~line intensities). 
Because of these charge exchange reactions, it is thus difficult to test 
stellar model atmospheres from the diagrams shown above, if nothing is known
about the ionization parameter. The only strong conclusion is that, for 
the Kurucz models to be consistent with the observations of nebulae with 
O$^{++}$/O$^+$ $<$ 1 or O$^{++}$/S$^{++}$ $<$ 10, one would require an 
unrealistically low ionization parameter, given the type of objects 
studied. On the other hand, atmosphere models including stellar winds, 
non--LTE effects and line blanketing naturally lead to approximately 
constant Ne$^{++}$/O$^{++}$ values over a large domain of stellar
temperatures in agreement with observations.

Another difficulty in the testing of model atmospheres by using H~{\sc ii}
regions is that many observed H~{\sc ii} regions are in fact complexes 
ionized by several O stars, up to hundreds of them. The best observational 
test would be in fact by modelling small single-star H~{\sc ii} regions 
for which integrated line emission strenghts would be available. 
To our knowledge, such data do not exist presently, but will be available 
soon (L. Deharveng, private communication).  

\section{Conclusions}

We have studied the impact of modern stellar atmospheres that take into 
account the effects of stellar winds, departures from LTE and line 
blanketing on the ionization structure of H~{\sc ii} regions.
Due to a flatter energy distribution in the He~{\sc i} continuum, 
compared to the widely used Kurucz models, generally higher ionic ratios 
are obtained, especially for ions with high ionization potentials. 
For He the effect is noticeable for stellar effective temperatures below 
$\sim$ 40000 K.

Using the {\em CoStar} model predictions we find that 
N$^{+}$/O$^{+}$ and Ne$^{++}$/O$^{++}$ can be safely used as direct
 indicators of N/O and Ne/O abundance ratios, 
over a wide range of effective temperatures and ionization parameters.

Comparisons of observed Ne$^{++}$/O$^{++}$ ionic ratios from IR line
measurements and optical lines in Galactic H~{\sc ii} regions 
(Shaver et al.~1983, Simpson et al.~1995) show that the {\em CoStar}
models naturally reproduce the roughly constant observed value of
Ne$^{++}$/O$^{++}$, while Kurucz models at solar metallicity fail
to reproduce this behaviour. This gives a strong support to predictions
of the ionizing fluxes from non--LTE atmosphere models including stellar
winds and line blanketing, in agreement with the results obtained
by Sellmaier et al.~(1996). 
Compared to observations of Magellanic Cloud H~{\sc ii} regions (Pagel
et al.~1978), the low metallicity Kurucz models, however, also show a 
reasonable agreement.

We also point out that tests of 
stellar atmosphere models from observations of H~{\sc ii} regions
may be inconclusive unless one has strong constraints on the ionization
parameter. 
Observations of integrated line strengths from single star H~{\sc ii} 
regions with well known exciting sources should allow one to set much
stronger constraints on the ionizing fluxes of O and B stars.
This would be extremely helpful to asses the reliability of current
models for O and B stars (cf. Cassinelli et al.~1995, Schaerer 1996).

\acknowledgements{DS acknowledges a fellowship from the Swiss National 
Foundation of Scientific Research and partial support from the 
Directors Discretionary Research Fund of the STScI. }




\begin{figure*}[htb]
\centerline{\psfig{figure=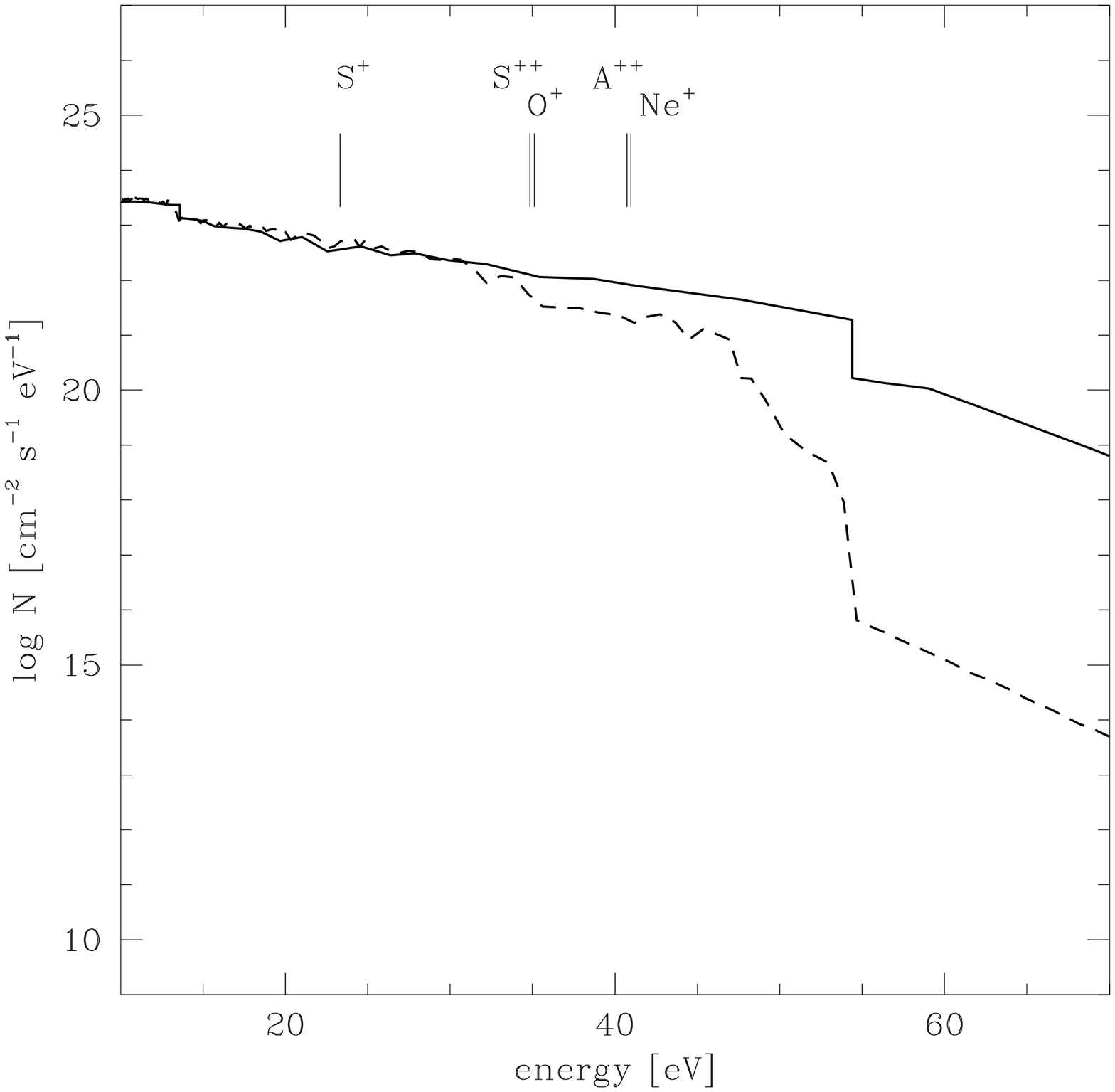,height=8.8cm}
            \psfig{figure=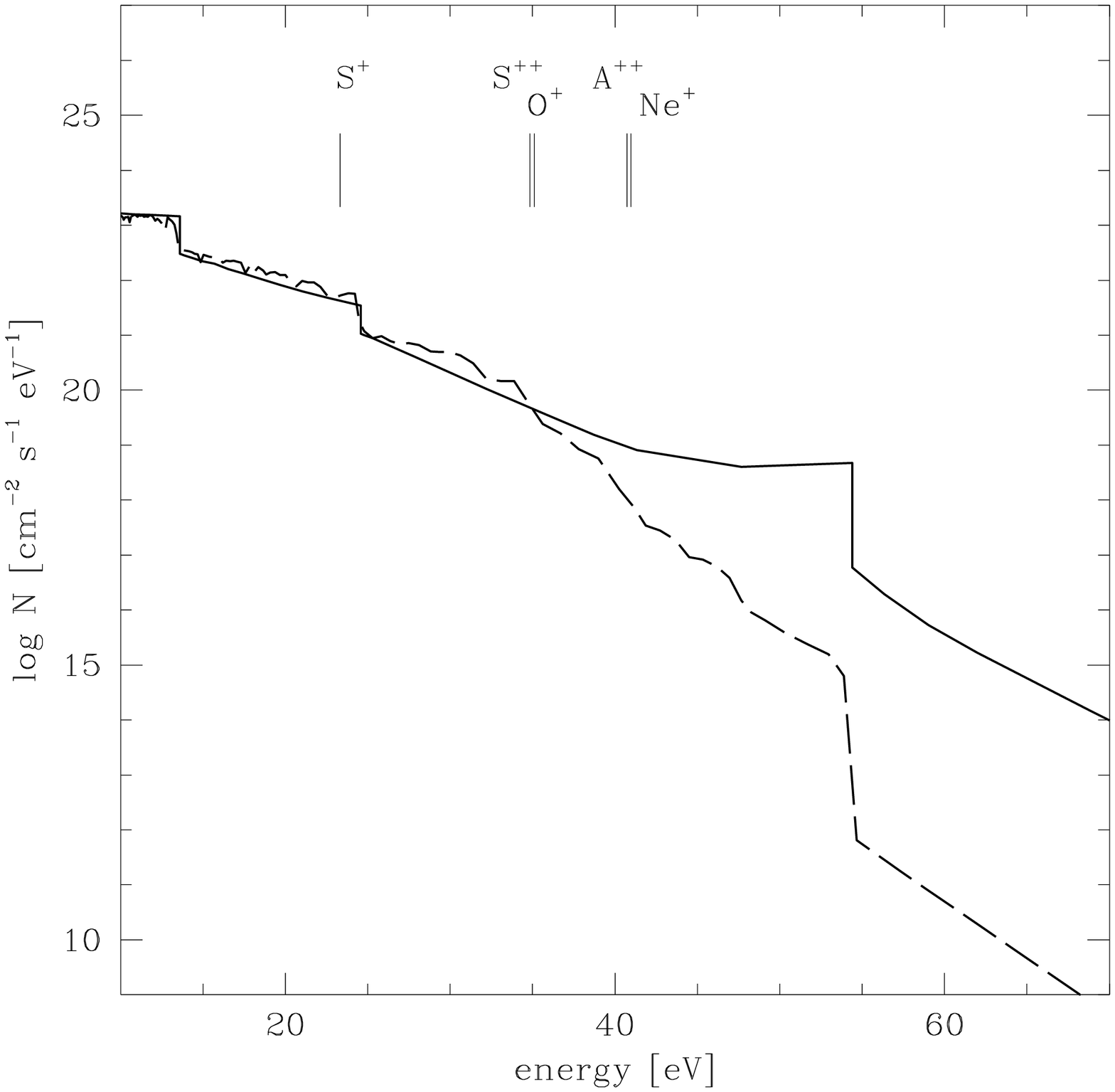,height=8.8cm}} 
\caption{Comparison of the EUV continous spectral energy distribution
 from {\em CoStar} models
 (solid lines) and Kurucz models (dashed). Plotted is the
 photon number flux per unit energy as a function 
 of the energy. Also shown are the ionization potentials of 
 $\rm S^{+}$, $\rm S^{++}$, $\rm O^{+}$, $\rm Ne^{+}$, and
 $\rm Ar^{++}$.
 {\em Left panel:} {\em CoStar} model D2: $(\teff,\logg)$ = 
 (46.1 kK, 4.05), Kurucz: $(\teff,\logg)$ = (45 kK, 5.0).
 {\em Right panel:}  {\em CoStar} model with $(\teff,\logg)$ 
= (35 kK, 4.0) and remaining parameters from model A1, 
Kurucz: $(\teff,\logg)$ =  (35 kK, 4.0) } 
\end{figure*}

\newpage
\begin{figure*}[htb]  
\centerline{
  \psfig{height=10.1cm,bbllx=1pt,bblly=4pt,bburx=750pt,bbury=527pt,clip=,figure=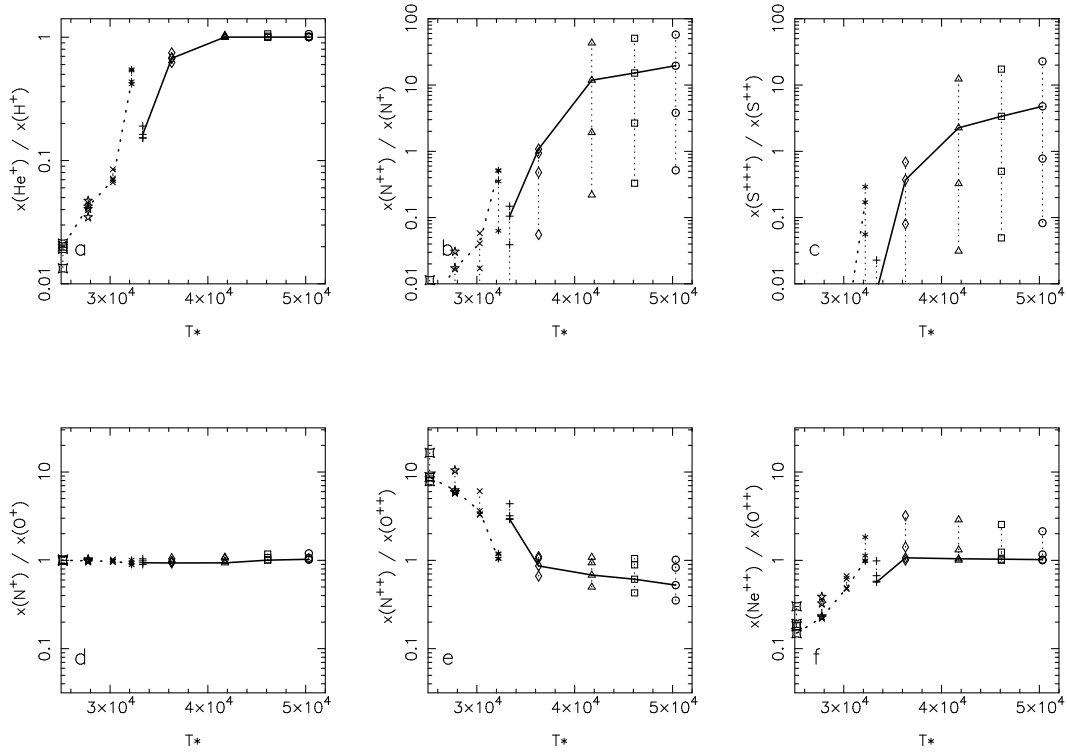}}
\caption{Predicted ionic fractions $x({\rm X^{+n}})/x({\rm Y^{+m}})$ 
as a function of the stellar effective temperature \teff\ from 
photoionization models using
{\em CoStar} atmospheres at solar metallicity (Z=0.020). 
At a given temperature the models 1 to 4, which differ by the ionization
parameter, are connected by a thin dotted line.
The ``reference models'' 2 are connected by the thick solid line for the
main sequence stars and by the thick dotted line for giants respectively.
See text for more explanations}
\end{figure*}

\begin{figure*}[htb] 
\centerline{
  \psfig{height=10.1cm,bbllx=1pt,bblly=4pt,bburx=750pt,bbury=527pt,clip=,figure=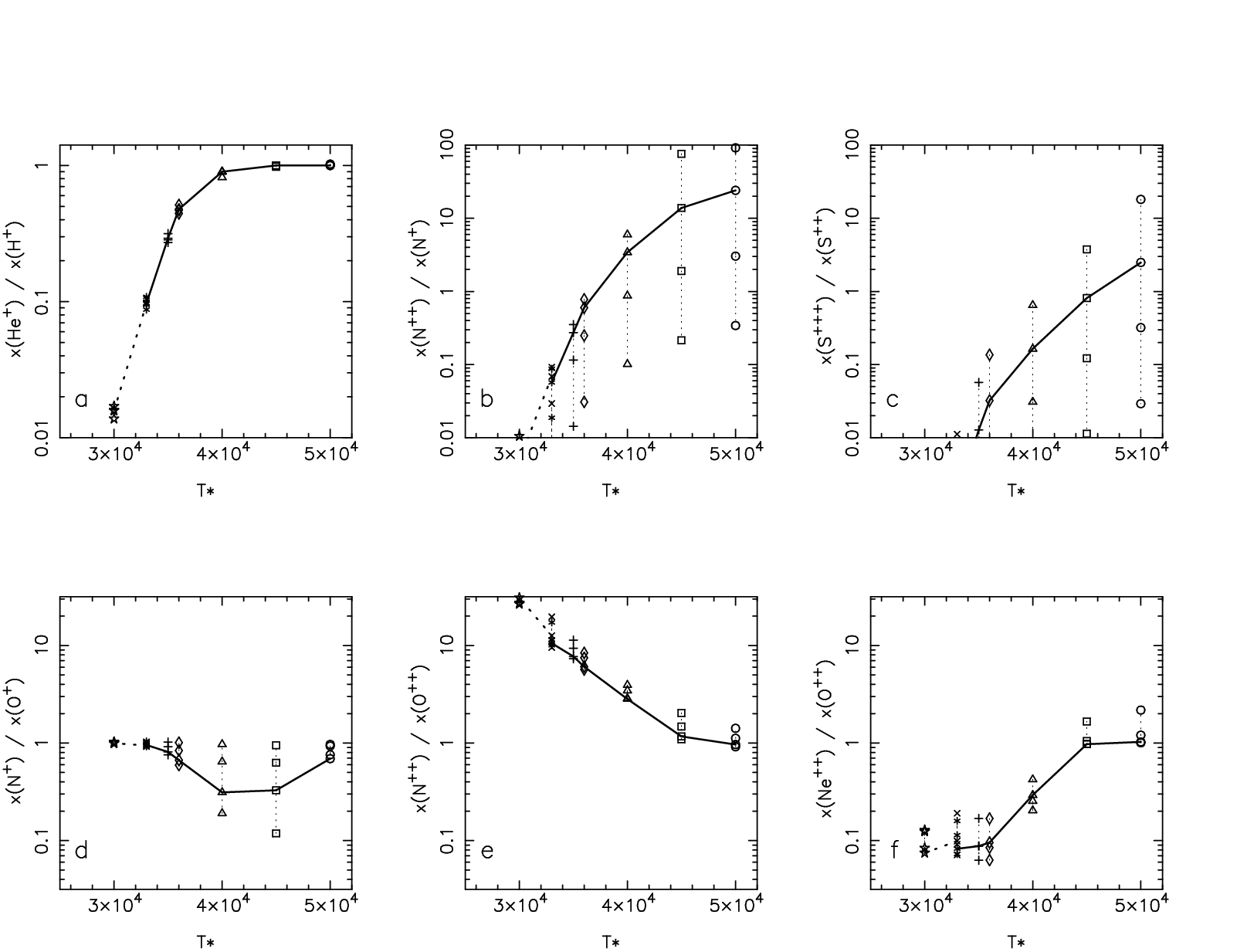}}
\caption{Same as Fig.~2 for the Kurucz (1991) models at solar metallicity}
\end{figure*}

\newpage
\begin{figure*}[htb] 
\centerline{
  \psfig{height=10.1cm,bbllx=1pt,bblly=4pt,bburx=750pt,bbury=527pt,clip=,figure=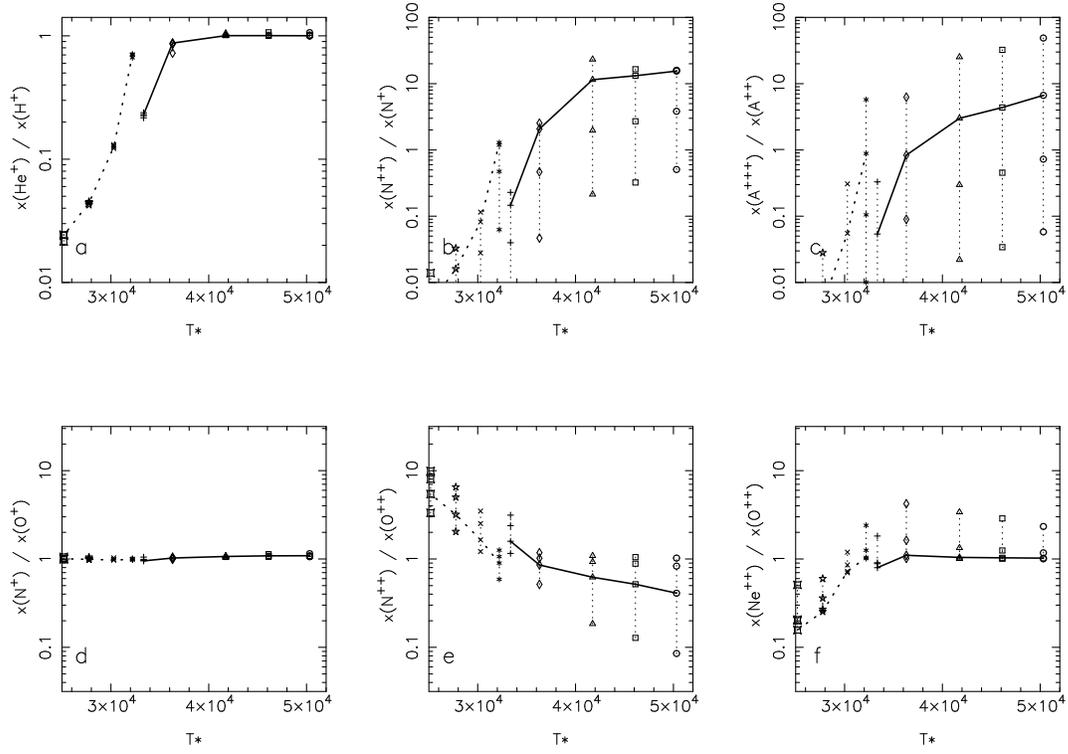}
}
\caption{Same as Fig.~2 for the low metallicity (Z=0.004) {\em CoStar} 
models}
\end{figure*}

\begin{figure*}[htb] 
\centerline{
  \psfig{height=10.1cm,bbllx=1pt,bblly=4pt,bburx=750pt,bbury=527pt,clip=,figure=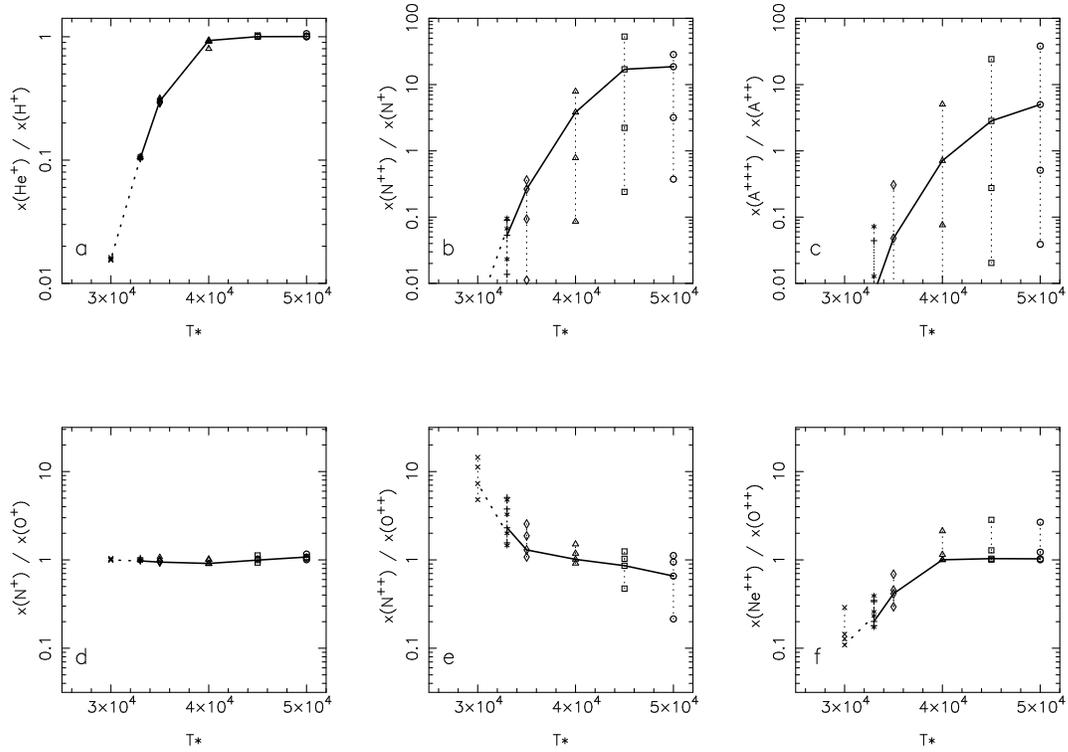}
}
\caption{Same as Fig.~2 for the low metallicity (Z=0.002) Kurucz (1991) 
models}
\end{figure*}

\clearpage
\begin{figure*}[htb] 
\centerline{
  \psfig{height=4.4cm,bbllx=70pt,bblly=18pt,bburx=300pt,bbury=550pt,clip=,figure=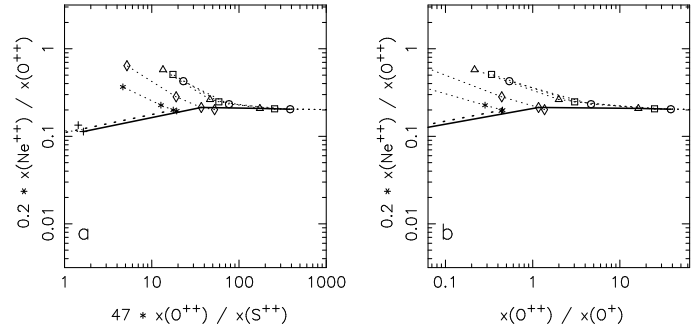,angle=270}
}
\caption{Predicted ionic Ne$^{++}$/O$^{++}$ abundance ratio as a function
of the O$^{++}$/S$^{++}$ ionic abundance ratio (panel {\bf a}) and
as a function of the O$^{++}$/O$^{+}$ ionic fraction (panel {\bf b})
for the solar metallicity {\em CoStar} models. Same symbols as in Fig.~2}
\end{figure*}

\begin{figure*} 
\centerline{
  \psfig{height=4.4cm,bbllx=70pt,bblly=40pt,bburx=300pt,bbury=550pt,clip=,figure=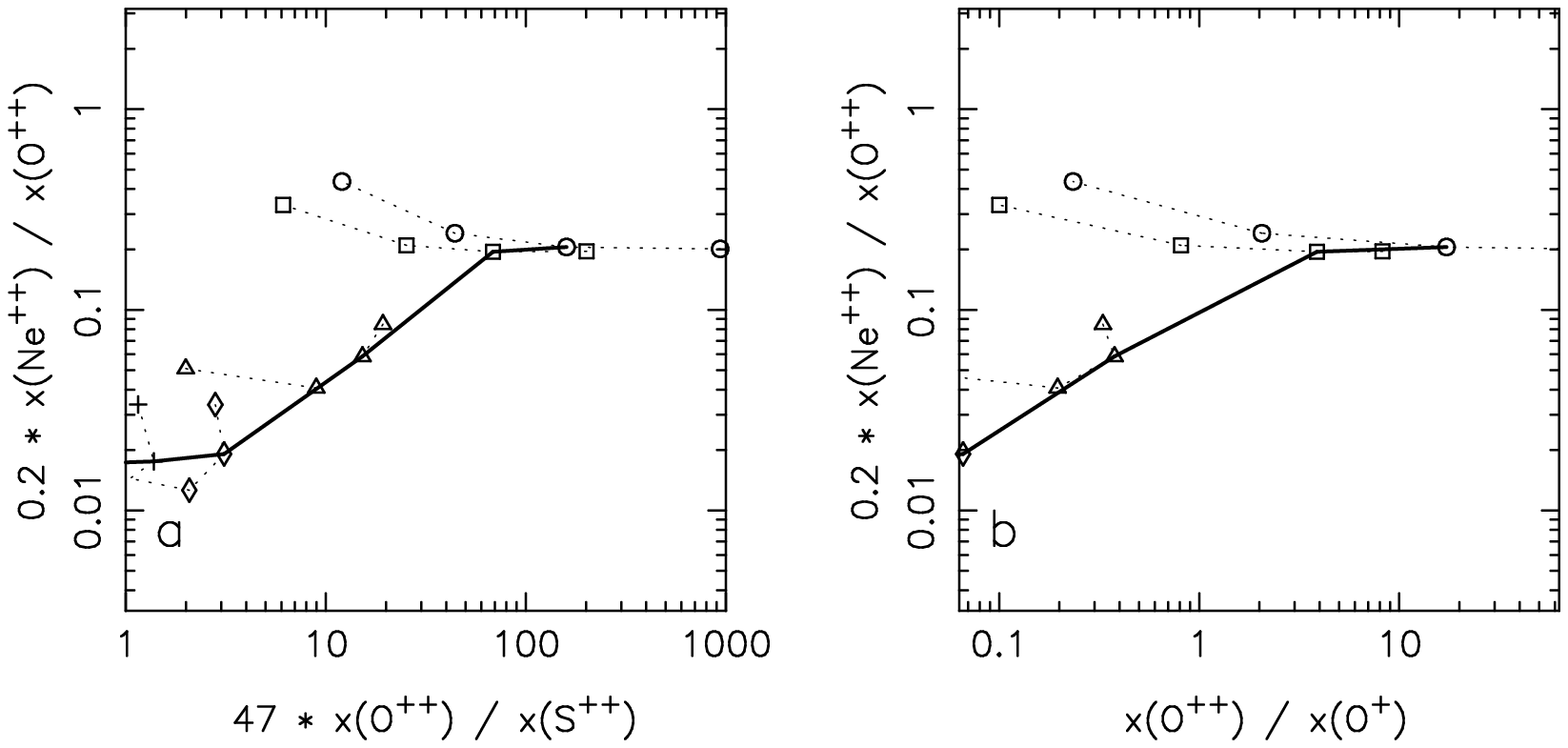,angle=270}
}
\caption{Same as Fig.~6 for the Kurucz (1991) models at solar metallicity}
\end{figure*}

\begin{figure*} 
\centerline{
  \psfig{height=4.4cm,bbllx=70pt,bblly=40pt,bburx=300pt,bbury=550pt,clip=,figure=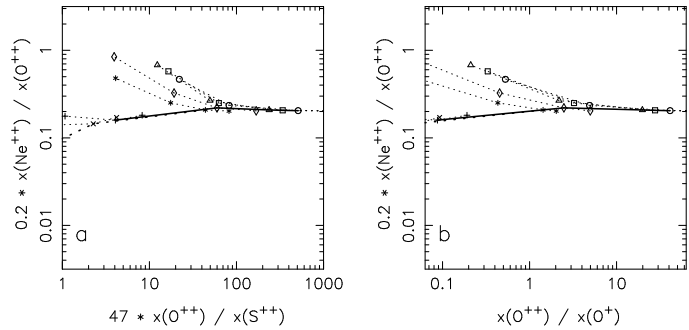,angle=270}
}
\caption{Same as Fig.~6 for the {\em CoStar} models at low metallicity
(Z=0.004)}
\end{figure*}

\begin{figure*} 
\centerline{
  \psfig{height=4.4cm,bbllx=70pt,bblly=18pt,bburx=300pt,bbury=550pt,clip=,figure=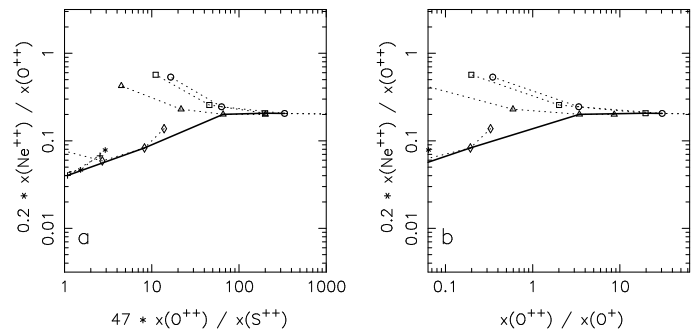,angle=270}
}
\caption{Same as Fig.~6 for the Kurucz (1991) models at low metallicity
(Z=0.002)}
\end{figure*}

\clearpage
\begin{figure}[htb]
\centerline{
  \psfig{height=4.65cm,bbllx=60pt,bblly=18pt,bburx=300pt,bbury=300pt,clip=,figure=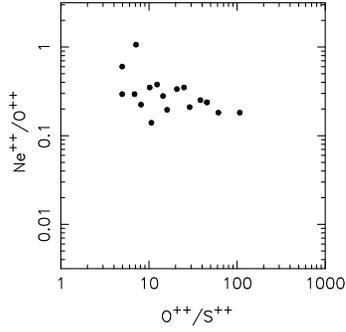,angle=270}}
\caption{Observed ionic Ne$^{++}$/O$^{++}$ abundance ratios as a function
of the O$^{++}$/S$^{++}$ ionic fractions in H~{\sc ii} regions. Data from
the IR line measurements
of Simpson et al.~(1995). To be compared with 
Figs.~6a -- 9 a}
\end{figure}

\begin{figure}[tb]
\centerline{
  \psfig{height=4.65cm,bbllx=60pt,bblly=18pt,bburx=300pt,bbury=300pt,clip=,figure=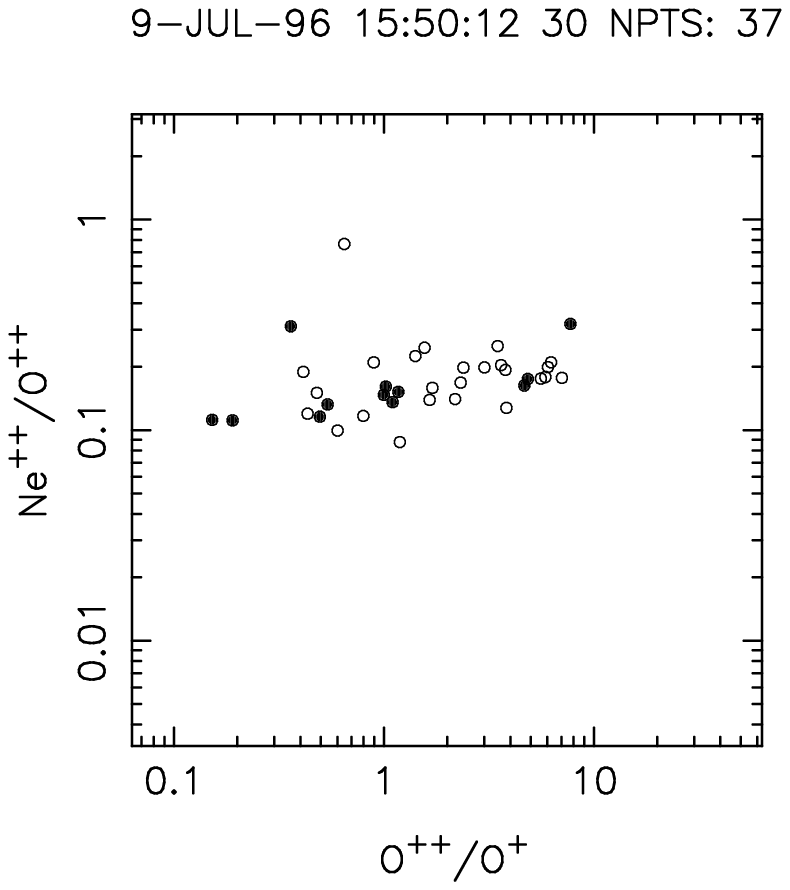,angle=270}}
\caption{Observed ionic Ne$^{++}$/O$^{++}$ abundance ratios as a function
of the O$^{++}$/O$^{+}$ ionic fractions. Filled circles: Galactic
H~{\sc ii} regions, data from Shaver et al.~(1983); open circles: Magellanic
Cloud H~{\sc ii} regions, data from Pagel et al.~(1978).
To be compared with Figs.~6b -- 9 b}
\end{figure}

\begin{figure}[b]
\centerline{
  \psfig{height=4.65cm,bbllx=60pt,bblly=18pt,bburx=300pt,bbury=300pt,clip=,figure=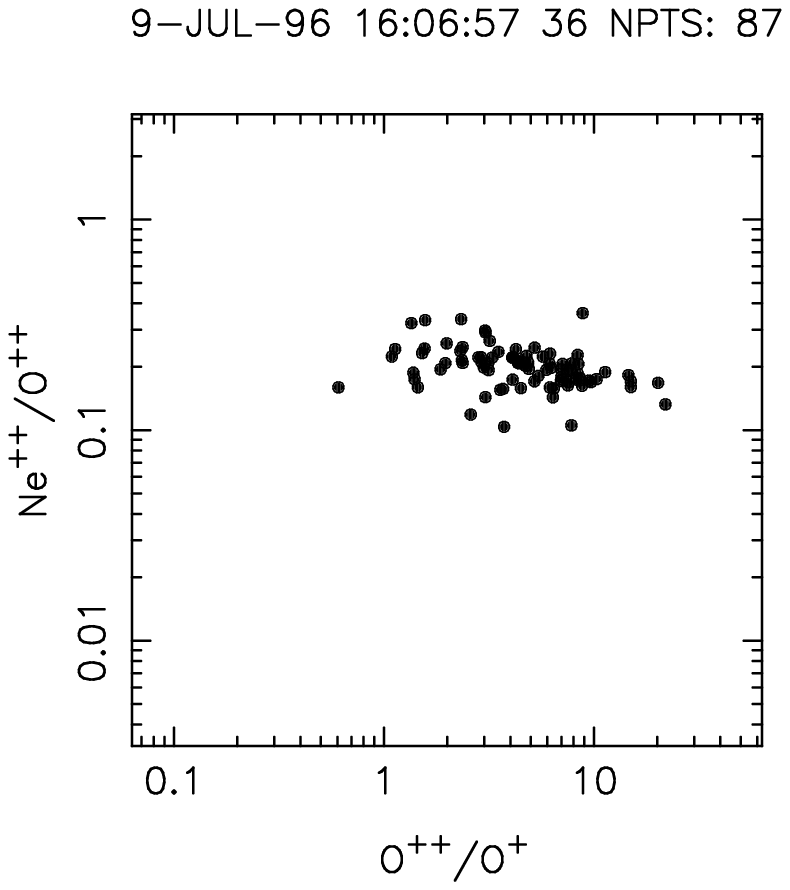,angle=270}}
\caption{Observed ionic Ne$^{++}$/O$^{++}$ abundance ratios as a function
of the O$^{++}$/O$^{+}$ ionic fractions. Data from the sample of
low metallicity H~{\sc ii} galaxies studied by SL96. 
To be compared with Figs.~6b -- 9 b}
\end{figure}

\end{document}